# Some Experimental Issues in Financial Fraud Detection: An Investigation

Jarrod West
School of Computing & Mathematics
Charles Sturt University
NSW, Australia-2640
jnwest@netspace.net.au

Maumita Bhattacharya
School of Computing & Mathematics
Charles Sturt University
NSW, Australia-2640
mbhattacharya@csu.edu.au

*Abstract*— Financial fraud detection is an important problem with a number of design aspects to consider. Issues such as algorithm selection and performance analysis will affect the perceived ability of proposed solutions, so for auditors and researchers to be able to sufficiently detect financial fraud it is necessary that these issues be thoroughly explored. In this paper we will revisit the key performance metrics used for financial fraud detection with a focus on credit card fraud, critiquing the prevailing ideas and offering our own understandings. There are many different performance metrics that have been employed in prior financial fraud detection research. We will analyse several of the popular metrics and compare their effectiveness at measuring the ability of detection mechanisms. We further investigated the performance of a range of computational intelligence techniques when applied to this problem domain, and explored the efficacy of several binary classification methods.

## I. Introduction and Background

Financial fraud is an increasingly prevalent issue that affects businesses and consumers alike worldwide and costs billions of dollars each year [1]. There are many different categories of financial fraud, including credit card fraud, financial statement fraud, securities and commodities fraud, insurance fraud, mortgage fraud, and money laundering [11]. An increase in online activity has led to a significant rise in reported financial fraud activity in recent years [1].

The simulations that we will present here explore several detection algorithms and metrics. Specifically, we will look at implementations of a number of computational intelligence techniques including variants of genetic algorithms, neural networks, and genetic programming solutions; as well as an ant colony optimisation, fuzzy logic, and a single statistical classification method (namely a support vector machine). Our analysis will continue on to review a number of binary classification metrics and how they fared at detecting financial fraud, concentrating specifically on credit card fraud.

Most of the prior research into feature selection and performance metrics has been included as part of specific experiments into detection algorithms. Fayyad studied the issues surrounding data mining and knowledge discovery in general, citing credit card fraud detection as a good example of the difficulties that arise with large and complicated problems [5]. Phua et al. performed an analysis on existing general fraud detection research [10]. As well as solution algorithms they investigated the performance metrics that researchers had used for various forms of fraud detection in the previous decade. Ngai et al. also looked at the distribution of prior research into various types of fraud and noticed that that there was an absence of study into visualisation and outlier methods [8]. Bhattacharyya et al. performed several credit card fraud experiments, comparing two common classification solutions against the well-known logistic regression and observing results across various common metrics [1]. Olszewski compared several forms of fraud detection and similar problems, including credit card fraud, telecommunications fraud, and network intrusion [9]. Dal Pozzolo et al. investigated several performance metrics and noted the importance of computational efficiency and cost minimisation [3]. West and Bhattacharya studied various forms of financial fraud as part of a broad review and concluded that there is a requirement for further research into the comparative performance of detection methods [11].

The rest of the paper is organized as follows: relevant performance metrics and associated issues are mentioned in Section II; simulation details and conclusions are presented in Section III and Section IV respectively.

## II. Performance Metrics and Relevant Issues

Measuring the success of computational intelligence algorithms is an important step in determining their suitability at solving their respective problem. This is especially true for a problem such as financial fraud, where minor improvements in performance can lead to large economic benefits. Performance can be measured in many different ways: absolute ability, performance relative to other factors, visual mediums, probability of success, and more. A list of some commonly used performance metrics for financial fraud detection is given in Table 1 [4], [6], [7].

## III. Simulations

To investigate the ability of classification techniques and efficacy of performance metrics we have run a number of experiments using binary classification methods and analysed the results. In particular we looked at accuracy, sensitivity, specificity, precision, false positive rate, F-measure, and a common variant of Fβ, F2. The following sections provide details on the various tests undertaken. We made use of 10-fold cross validation to reduce the chance of statistical errors.

TABLE I. COMMONLY USED PERFORMANCE METRICS FOR SEVERAL PROBLEM REPRESENTATIONS

| Problem | Metric | Description | Formula |
|---|---|---|---|
| Classification | Accuracy | The ratio of samples correctly classified to total samples | $ACC = (TP+TN)/(P+N)$ |
| | Sensitivity | The ratio of positive samples correctly classified to total positive samples. Also known as recall, true positive hit rate, or hit rate | $SENS = TP/P$ |
| | Specificity | The ratio of negative samples correctly classified to total negative samples | $SPEC = TN/N$ |
| | Precision | The ratio of positive samples correctly classified to total samples classified as positive | $PREC = TP/(TP+FP)$ |
| | False positive rate | The inverse of the true positive rate (sensitivity) | $FPR = FP/N$ |
| | F-measure | Also known as F-score or F, the F-measure is the harmonic mean of precision and recall (sensitivity) | $F = (2 \times PREC \times SENS)/(PREC+SENS)$ |
| | Fβ | A form of F-measure that applies a weighting of *β* to the precision and recall | $F_\beta = ((1+\beta^2) \times PREC \times SENS)/(\beta^2 \times PREC+SENS)$ |
| | Cost minimisation | An algorithm that minimizes cost associated with each type of error | $C = FPR \times CFP + FNR \times CFN$ |
| Statistical | Z-score | Measures the rate of change in a variable, either independently, with respect to its historical values, or against a similar variable | $Z = (X-\mu)/\sigma$ |
| | Sum of squared error | A measurement of the difference between two sets of values, squared to separate out distinct clusters of values: | $E = \sum_{i=1}(y_{Ei} - y_{Ai})^2$ |
| Association rule | Support | Calculates the percentage of samples that contain a specific itemset | $S_I = (\sum X_{Ij})/N$ |
| | Confidence | A measure of the effectiveness of a rule using the support of samples that conform to the rule | $C_{X \to Y} = S_{X \cup Y}/S_X$ |
| | Lift | A correlation method used to determine whether an association rule is useful to the problem | $L_{X \to Y} = C_{X \to Y}/S_Y$ |
| | Conviction | A measure of the inaccuracy of a rule | $CV_{X \to Y} = (1-S_Y)/(1-C_{X \to Y})$ |
| Clustering | Hopkins statistic | The probability that a variable is randomly distributed within a space, used to determine whether a problem contains significant clusters | $H = (\sum_{i=1} y_i)/(\sum_{i=1} x_i + \sum_{i=1} y_i)$ |
| Visualisation | AUC | The area under a Receiver Operating Characteristic (ROC) curve, provides a numeric value for a visualization problem using true and false positive hit rates | $AUC = \int ROC$ |

*A. Algorithms*

- *GP1* – A hybrid of two common genetic programming approaches, Pittsburgh and Michigan.
- *GP2* – An approach that represents individuals as a single rule with a context free grammar.
- *GA1* – An incrementally learning algorithm that assesses attributes individually and orders them based on their relevance to the problem.

- *GA2* – An extended version of GA1 that can use either the best rule as the basis of the next generation or the entire ruleset.
- *ACO* – A model of an ant colony with rules representing the path that an individual ant will follow.
- *NN1* – An incrementally created neural network that uses distance weighting to construct its hidden neurons.
- *NN2* – An iterative network that represents its neurons as learning vectors.
- *SVM* – A slight variant of a typical support vector machine with parameterised support vectors.
- *FL* – A fuzzy rule learner based on the RIPPER algorithm.

*B. Dataset*

The UCSD-FICO Data mining contest 2009 credit card dataset used in this re-search consists of 10000 samples, 334 input attributes which are all numerical, and two output classes for fraudulent and legitimate behaviour. Like other fraud datasets the UCSD dataset is comprised of severely unbalanced data, with only 9% representing fraudulent transactions.

*C. Results and Analysis*

Numerical results for the experiments are given in Table 2. We've reported values for all the binary classification metrics listed in Table 1, with the exception of cost minimisation as there was no readily available information on the costs associated with output classes in the dataset. Additionally, we've listed the percentage of classification error in Table 3, defined as the number of false positives and false negatives as a percentage of the total samples. Finally we've included the standard deviation of each metric across all cross validation folds in Table 4.

With the aid of the metrics we can immediately see that several algorithms achieved a very high performance across different criteria. The SVM, GP2, FL, and both GA1 and GA2 had accuracies of greater than 90% and false positive rates of less than 0.002 between them. The SVM algorithm in particular accomplished the highest accuracy, specificity, and precision, as well as the lowest false positive rate and reasonable results for the remaining metrics.

Despite their overall success it can also be observed that each algorithm had a significantly inferior specificity, as low as 0.8% in the case of GA1. The most likely reason for this is due to overtraining given the vast imbalance between classes: the UCSD dataset has a ratio of approximately 9:91 fraudulent to legitimate transactions. However, there is a correlation that can be drawn between sensitivity and accuracy: the algorithms with lower accuracy also had a higher sensitivity. This can be observed with both of the neural networks variants as well as GP1. A potential cause of this is the incremental nature of both neural networks resulting in the positive values to be given larger consideration than other algorithms, as well as GP1s fitness function determining generational success using both sensitivity and specificity [2].

Although several methods had lower accuracy their tendency to a higher sensitivity than other algorithms may result in them having better performance when considering cost minimisation. A sufficiently large ratio between the cost of false positive and false negative, which is frequently present in many forms of fraud detection, may result in these methods being more suitable to specific problems. This emphasises the importance of understanding and determining the problem requirements before beginning the experiments so the correct metrics can be chosen to evaluate performance.

Fig 1 shows the percentage of classification error for each validation fold used on the dataset. Most of the algorithms showed considerable consistency between each fold, however both neural network implementations and the ant colony optimisation had a noticeable amount of variance. This may be due to the iterative nature of these algorithms, which leads to a higher possibility of divergence at the start of their execution. Further differences between each fold have been identified in Table 3, which shows the standard deviation of each metric on every method.

TABLE II. POPULAR BINARY CLASSIFICATION PERFORMANCE METRICS FOR RESULTS ON THE DATASET

| Algorithm | Accuracy | Sensitivity | Specificity | Precision | False positive rate | F-measure | $F_2$ |
|---|---|---|---|---|---|---|---|
| GP1 | 0.606 | 0.493 | 0.618 | 0.114 | 0.382 | 0.185 | 0.296 |
| GP2 | 0.910 | 0.025 | 0.998 | 0.548 | 0.002 | 0.049 | 0.031 |
| GA1 | 0.910 | 0.008 | 1.000 | 0.636 | 0.000 | 0.015 | 0.010 |
| GA2 | 0.911 | 0.016 | 1.000 | 0.778 | 0.000 | 0.031 | 0.019 |
| ACO | 0.892 | 0.032 | 0.978 | 0.125 | 0.022 | 0.051 | 0.038 |
| NN1 | 0.771 | 0.271 | 0.821 | 0.131 | 0.179 | 0.176 | 0.223 |
| NN2 | 0.774 | 0.193 | 0.832 | 0.103 | 0.168 | 0.134 | 0.164 |
| SVM | 0.915 | 0.064 | 1.000 | 0.951 | 0.000 | 0.120 | 0.079 |

TABLE III. PERCENTAGE OF CLASSIFICATION ERROR FOR EACH METHOD PER FOLD

| Algorithm | GP1 | GP2 | GA1 | GA2 | ACO | NN1 | NN2 | SVM | FL |
|---|---|---|---|---|---|---|---|---|---|
| Fold 1 | 0.361 | 0.088 | 0.090 | 0.088 | 0.080 | 0.254 | 0.155 | 0.081 | 0.089 |
| Fold 2 | 0.397 | 0.090 | 0.088 | 0.087 | 0.090 | 0.250 | 0.283 | 0.086 | 0.088 |
| Fold 3 | 0.385 | 0.090 | 0.090 | 0.087 | 0.089 | 0.383 | 0.092 | 0.087 | 0.089 |
| Fold 4 | 0.383 | 0.092 | 0.090 | 0.088 | 0.090 | 0.244 | 0.183 | 0.087 | 0.091 |
| Fold 5 | 0.393 | 0.092 | 0.091 | 0.091 | 0.091 | 0.188 | 0.246 | 0.085 | 0.093 |
| Fold 6 | 0.414 | 0.089 | 0.090 | 0.091 | 0.109 | 0.202 | 0.346 | 0.086 | 0.091 |
| Fold 7 | 0.399 | 0.091 | 0.089 | 0.087 | 0.097 | 0.229 | 0.261 | 0.083 | 0.093 |
| Fold 8 | 0.412 | 0.091 | 0.091 | 0.091 | 0.141 | 0.256 | 0.201 | 0.083 | 0.091 |
| Fold 9 | 0.393 | 0.090 | 0.091 | 0.090 | 0.116 | 0.240 | 0.230 | 0.085 | 0.092 |
| Fold 10 | 0.399 | 0.088 | 0.091 | 0.088 | 0.159 | 0.202 | 0.159 | 0.087 | 0.091 |

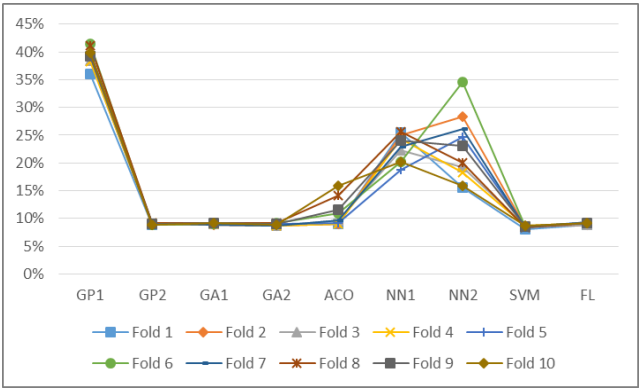

Fig. 1. Percentage of classification error for each method per fold

TABLE IV. STANDARD DEVIATION OF PERFORMANCE METRICS PER ALGORITHM OVER EACH FOLD

| Algorithm | Accuracy | Sensitivity | Specificity | Precision | False positive rate | F-measure | F2 |
|---|---|---|---|---|---|---|---|
| GP1 | 0.014 | 0.047 | 0.015 | 0.010 | 0.015 | 0.016 | 0.026 |
| GP2 | 0.001 | 0.022 | 0.003 | 0.247 | 0.003 | 0.030 | 0.021 |
| GA1 | 0.001 | 0.012 | 0.001 | 0.415 | 0.001 | 0.009 | 0.006 |
| GA2 | 0.002 | 0.013 | 0.001 | 0.200 | 0.001 | 0.016 | 0.011 |
| ACO | 0.023 | 0.033 | 0.027 | 0.152 | 0.027 | 0.029 | 0.027 |
| NN1 | 0.023 | 0.038 | 0.027 | 0.016 | 0.027 | 0.018 | 0.025 |
| NN2 | 0.057 | 0.094 | 0.071 | 0.023 | 0.071 | 0.030 | 0.053 |
| SVM | 0.002 | 0.022 | 0.001 | 0.075 | 0.001 | 0.039 | 0.027 |
| FL | 0.002 | 0.012 | 0.002 | 0.273 | 0.002 | 0.020 | 0.013 |

Because each fold is a subset of the same dataset large variances in the results between folds can indicate a higher sensitivity to minor changes in the data, which is a desirable trait in a complicated problem like financial fraud detection. Precision showed the greatest standard deviation out of any metric, most likely as another consequence of the imbalanced nature of the dataset resulting in a larger variance for the metrics that analyses them. As this imbalance is typical for the financial fraud detection problem precision is likely an excellent metric for assessing algorithm performance.

Additionally we can see that the ant colony optimisation and both neural networks all tended to a higher standard deviation across metrics than other methods. Specifically NN2 had the highest out of all methods, achieving 0.094 on sensitivity and 0.071 on specificity. These results shows a correlation with the higher variance in error rates found between these three methods earlier, and supports our claim that the iterative structure of these algorithms leads to the chance of small variance during the initial stages compounding into larger differences by completion. The receptiveness of the ant colony optimisation and neural network methods to small differences between each fold makes them suitable algorithms for financial fraud detection.

## IV. CONCLUSION

This paper analysed some key experimental issues associated with financial fraud detection, namely the ability of detection algorithms and the metrics used to assess their performance focusing on credit card fraud detection. We conducted several experiments, using various computational intelligence algorithms and analysing the results using a number of binary classification metrics. We found that our GP1 (a genetic programming variant) implementation and both neural networks showed superior results for specificity, an important metric given the imbalanced misclassification costs involved with fraud. The neural networks and ant colony optimisation also showed a higher receptiveness to changes in the dataset which is another desirable trait. Finally, we discovered that precision showed the highest variance between performance metrics, suggesting that it may be a good choice for assessing financial fraud detection solutions. Of course, we acknowledge that choice of performance metric often depends on the fraud detection goals.

There are other aspects of financial fraud detection that would warrant further investigation, and additional research could explore these issues in greater depth, particularly with a focus on specific detection methods. Also, future researchers may wish to concentrate on additional comparisons between the various performance metrics using further controlled experiments.